\documentclass[a4paper]{amsart}
\usepackage{amscd}
\usepackage{amsmath}
\usepackage{amssymb}
\usepackage{amsthm}
\usepackage{bbm} 
\usepackage{stmaryrd}

\usepackage[T1]{fontenc}

\newif\ifpdf
\ifx\pdfoutput\undefined
   \pdffalse        
\else
   \pdfoutput=1     
   \pdftrue
\fi

\ifpdf
   \usepackage[pdftex]{graphicx}
\else
   \usepackage{graphicx}
\fi

\numberwithin{equation}{section} \swapnumbers

\newcommand{\bbr}{\mathbb{R}}

\begin{document}

\title[A simple mathematical model for the corona virus]{A simple mathematical model for the evolution of the corona virus}
\author{Stefan Tappe}
\address{Ludwig Maximilian University of Munich, Department of Mathematics, Theresienstr. 39, 80333 Munich, Germany}
\email{tappe@math.lmu.de}
\date{20 March, 2020}
\begin{abstract}
The goal of this note is to present a simple mathematical model with two parameters for the number of deaths due to the corona (COVID-19) virus. The model only requires basic knowledge in differential calculus, and can also be understood by pupils attending secondary school. The model can easily be implemented on a computer, and we will illustrate it on the basis of case studies for different countries.
\end{abstract}

\maketitle\thispagestyle{empty}

\section{Introduction}

The corona (COVID-19) virus is currently a challenge for several countries of the world. There are several recent articles, such as \cite{Liu, Roosa, Tang, Tang-2, Thompson, Wu}, which provide mathematical models. The goal of this note is to present a rather simple mathematical model for the evolution of the corona virus, where we merely concentrate on the number of deaths due to the corona virus. However, from adequate predictions of these numbers of deaths, we will also obtain estimates for the number of infected people and for the probability that an individual of a population is infected at a certain time point. The mathematical model is intended to be simple enough such that practitioners can use it in order to compute different scenarios for the future. Even pupils attending secondary school with basic knowledge in differential calculus should be able to understand the model.

The remainder of this note is organized as follows. In Section \ref{sec-motivation} we have a look at the data from China, and use this data in order to motivate the mathematical properties of our model. Afterwards, in Section \ref{sec-presentation} we present the mathematical model. In Section \ref{sec-China} we perform the fitting procedure of the parameters for our model using the data from China. In Section \ref{sec-Italy} we look at the data from Italy and make predictions for the number of deaths according to our model, and make some concluding remarks regarding the situation in other countries. All data which we use in this note are taken from \cite{Worldometer}, and all plots in this note are generated using R.

\section{A look at the data from China}\label{sec-motivation}

\begin{figure}[!ht]
 \centering
 \includegraphics[width=0.4\textwidth]{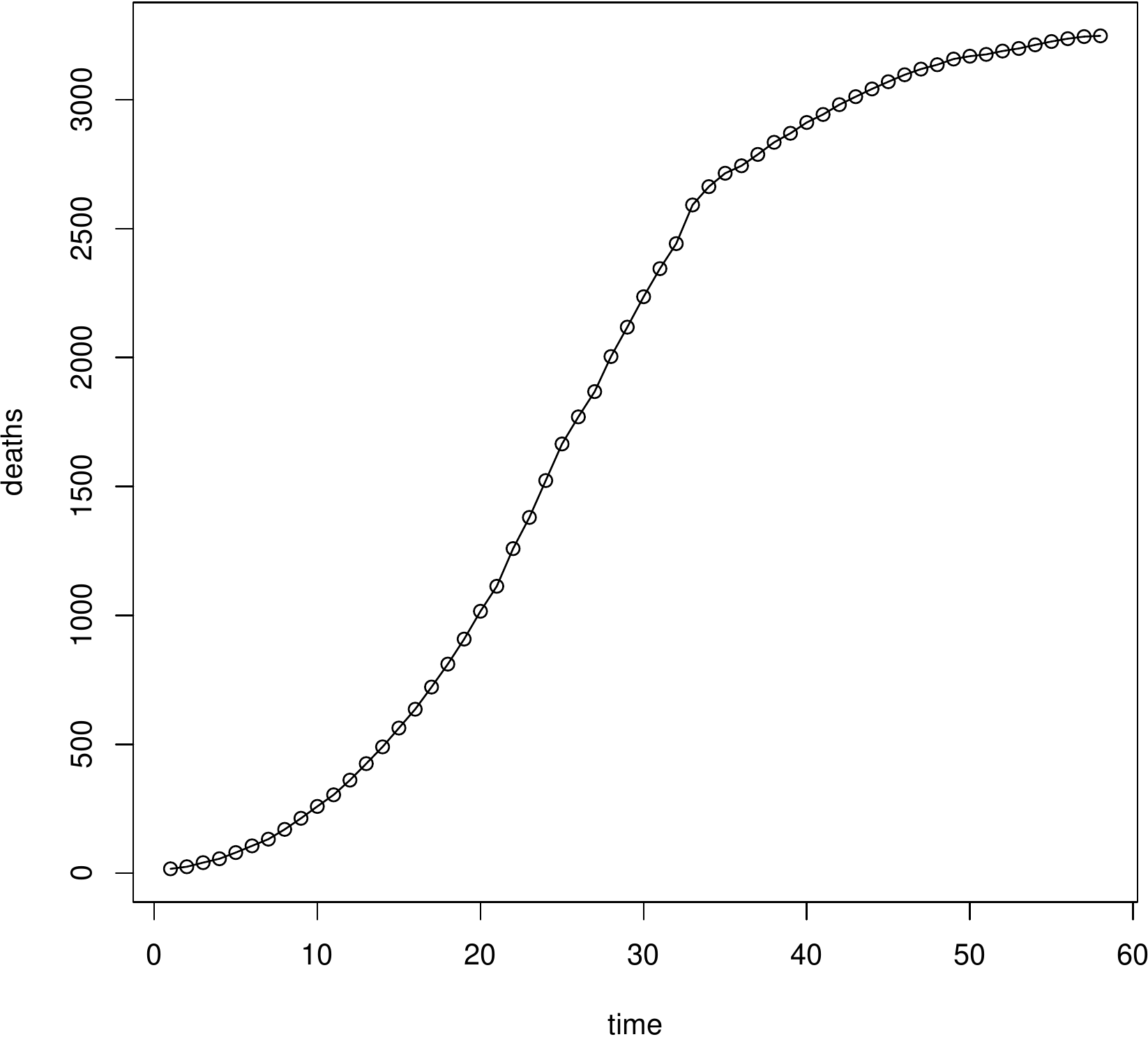}
 \includegraphics[width=0.4\textwidth]{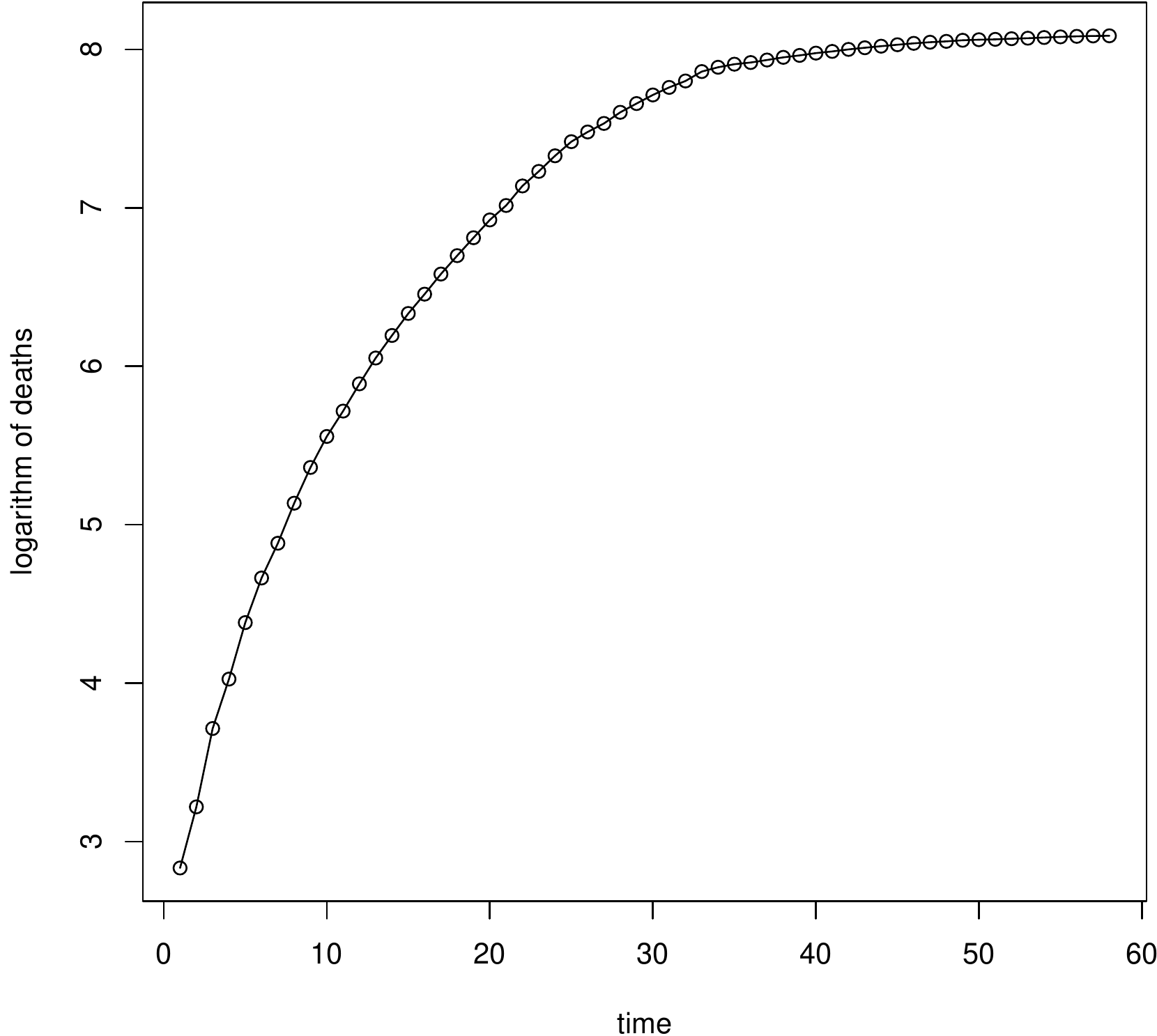}
 \caption{These two plots show the numbers of deaths in China from January 22 to March 19; in the left plot with the usual linear scale, and in the right plot with logarithmic scale.}\label{fig-china}
\end{figure}

In order to motivate our model, let us perform an empirical investigation of Chinese data. So far, China is the only country which has essentially already overcome the corona crisis, and where the numbers of deaths due to the corona virus are stabilizing. Figure \ref{fig-china} shows the numbers of deaths in China from January 22 to March 19, with linear and logarithmic scale.

Although these data are only available on a daily basis, we assume that the number of deaths is a function $D : [t_1,\infty) \to [1,\infty)$ defined on an interval, where $t_1$ is the first day when deaths due to the corona virus are observed. Then we can write the number of deaths as
\begin{align*}
D(t) = \exp( L(t) ), \quad t \in [t_1,\infty),
\end{align*}
where $L : [t_1,\infty) \to \bbr_+$ is the logarithm. With this notation, the function $D$ appears in the left plot of Figure \ref{fig-china}, and the function $L$ appears in the right plot of Figure \ref{fig-china}. Having a closer look at Figure \ref{fig-china}, we make the following observations:
\begin{enumerate}
\item The function $L$ is strictly increasing and continuous, in fact even continuously differentiable. As a consequence, the function $D$ is strictly increasing and continuously differentiable as well.

\item The function $L$ is concave on the whole domain $[t_1,\infty)$.

\item There is time point $T_1$ at which the curvature of $D$ changes. Namely, the function $D$ is convex on $[t_1,T_1]$ and concave on $[T_1,\infty)$. The time point $T_1$ is the time where governmental measures start to become effective. Since an infected person who dies, will die around $17$ days after infection, in our model the time point $T_1$ is $17$ days after the governmental measures.

\item The limit $L(\infty) := \lim_{t \to \infty} L(t)$ exists. As a consequence, the limit $D(\infty) := \lim_{t \to \infty} D(t)$ exists as well. This means that the number of deaths stabilizes, and indicates that the corona crisis is essentially overcome.
\end{enumerate}

\section{Presentation of the model}\label{sec-presentation}

Bases on the empirical observations from Section \ref{sec-motivation}, we will now present our model for the number of deaths. As suggested in the previous section, we consider a function $D : [t_1,\infty) \to [1,\infty)$. Typically we choose $t_1 = 1$, the first day when deaths due to the corona virus are observed. Then we can write the number of deaths as
\begin{align*}
D(t) = \exp( L(t) ), \quad t \in [t_1,\infty),
\end{align*}
where $L : [t_1,\infty) \to \bbr_+$ is the logarithm. Let $T_1 \in [t_1,\infty)$ be the time point where governmental measures begin to show consequences. As already discussed, this is typically $17$ days after the governmental measures. Let $(t_1,\ldots,t_n)$ be a vector of time points for which the numbers of deaths $(d_1,\ldots,d_n)$ are available from data. We assume that $t_n \leq T_1$, and denote by $(l_1,\ldots,l_n)$ the corresponding logarithms
\begin{align*}
l_i := \ln(d_i), \quad i=1,\ldots,n.
\end{align*}
Now, we introduce the function $L : [t_1,\infty) \to \bbr_+$ as follows:
\begin{enumerate}
\item On the interval $[t_1,T_1]$ we define
\begin{align*}
L(t) := l_1 + (l_n - l_1) \cdot \bigg( \frac{t-t_1}{t_n-t_1} \bigg)^{\beta}, \quad t \in [t_1,T_1]
\end{align*}
for some constant $\beta \in (0,1]$, the concavity parameter. Then we have $L(t_1)=l_1$ and $L(t_n)=l_n$, which means that the function $L$ matches with the data points at $t_1$ and $t_n$.

\item On the interval $(T_1,\infty)$ we define 
\begin{align*}
L(t) := L(T_1) + \lambda \big( 1 - \exp ( -\nu (t-T_1) ) \big), \quad t \in (T_1,\infty)
\end{align*}
with parameters $\lambda,\nu > 0$.
\end{enumerate}
Then the following statements are true:
\begin{enumerate}
\item The function $L$ is strictly increasing and continuous.

\item Computing the second order derivative of $L$, we see that the function $L$ is concave.

\item Computing the second order derivative of $D$, we see that $D$ is convex on $[t_1,T_1]$ and concave on $[T_1,\infty)$.

\item The limit $L(\infty) := \lim_{t \to \infty} L(t)$ exists with
\begin{align*}
L(\infty) = L(T_1) + \lambda.
\end{align*}
Therefore, the limit $D(\infty) := \lim_{t \to \infty} D(t)$ exists as well, and we have
\begin{align*}
D(\infty) = D(T_1) \cdot \exp(\lambda).
\end{align*}
\end{enumerate}
The only missing property is that so far the function $L$ does not need to be continuously differentiable at $T_1$. For this, we require conditions on the parameters $\lambda$ and $\nu$. We have
\begin{align*}
L'(t) = \beta \cdot \frac{l_n - l_1}{t_n-t_1} \cdot \bigg( \frac{t-t_1}{t_n-t_1} \bigg)^{\beta-1}, \quad t \in [t_1,T_1),
\end{align*}
and the left-hand derivative at $T_1$ is given by
\begin{align*}
\lim_{h \downarrow 0}\frac{L(T_1)-L(T_1-h)}{h}  = \beta \cdot \frac{l_n - l_1}{t_n-t_1} \cdot \bigg( \frac{T_1-t_1}{t_n-t_1} \bigg)^{\beta-1}.
\end{align*}
Moreover, we have
\begin{align*}
L'(t) = \lambda \nu \cdot \exp ( -\nu (t-T_1) ), \quad t \in (T_1,\infty), 
\end{align*}
and the right-hand derivative at $T_1$ is given by
\begin{align*}
\lim_{h \downarrow 0}\frac{L(T_1+h)-L(T_1)}{h}  = \lambda \nu.
\end{align*}
Therefore, the function $L$ is continuously differentiable at $T_1$ if and only if
\begin{align}\label{nu}
\nu = \frac{\beta}{\lambda} \cdot \frac{l_n - l_1}{t_n-t_1} \cdot \bigg( \frac{T_1-t_1}{t_n-t_1} \bigg)^{\beta-1}.
\end{align}
Consequently, in our model we only specify the two parameters $(\beta,\lambda)$, and then the parameter $\nu$ is already determined by (\ref{nu}). The two parameters $(\beta,\lambda)$ have the following interpretations:
\begin{itemize}
\item The parameter $\beta \in (0,1]$ is the concavity parameter of $L$. If $\beta = 1$, then the function $L$ is linear. Otherwise, the function $L$ is concave, and decreasing the value of $\beta$ results in increasing the concavity of the function $L$.

\item The parameter $\lambda > 0$ determines how rigorous the governmental measures are. As we have seen, the asymptotic number of deaths is given by
\begin{align*}
D(\infty) = D(T_1) \cdot \exp(\lambda).
\end{align*}
Hence, small values of $\lambda$ indicate rigorous governmental measures, whereas large values of $\lambda$ indicate only gentle governmental measures, with the consequence of a larger number of deaths in the long run.
\end{itemize}
So far, we have focused on the number of deaths $D(t)$ due to the corona virus at time $t$. Another quantity of interest is the number $I(t)$ of people infected with the corona virus at time $t$. It is well known that the official numbers are not realistic. They are much too low, because the government is only aware of reported cases. However, given the function $D : [t_1,\infty) \to [1,\infty)$ of the numbers of deaths, an estimate for the number of infected people at time $t$ is given by
\begin{align*}
I(t) = \frac{D(t+17)}{\mu \cdot \kappa}.
\end{align*}
Here $\mu \in (0,1)$ is the mortality rate of infected people, and $\kappa > 0$ is a constant due to cluster effects. Reasonable values are, for example $\mu = 0.01$ and $\kappa = 5$. We refer, for example, to the discussion in \cite{Pueyo}. Considering the fractions
\begin{align*}
P(t) = \frac{I(t)}{n},
\end{align*}
where $n$ denotes the size of the population, we also obtain estimates for the probabilities that at time $t$ an individual is infected with the corona virus.

\section{Estimates for Chinese data}\label{sec-China}

In this section we come back to the data from China, and fit the parameters $(\beta,\lambda)$ in our model. The concavity parameter is estimated as $\beta = 0.6$, and the parameter representing the rigorousness of governmental measures is estimated as $\lambda = \ln(3.25)$.

\begin{figure}[!ht]
 \centering
 \includegraphics[width=0.5\textwidth]{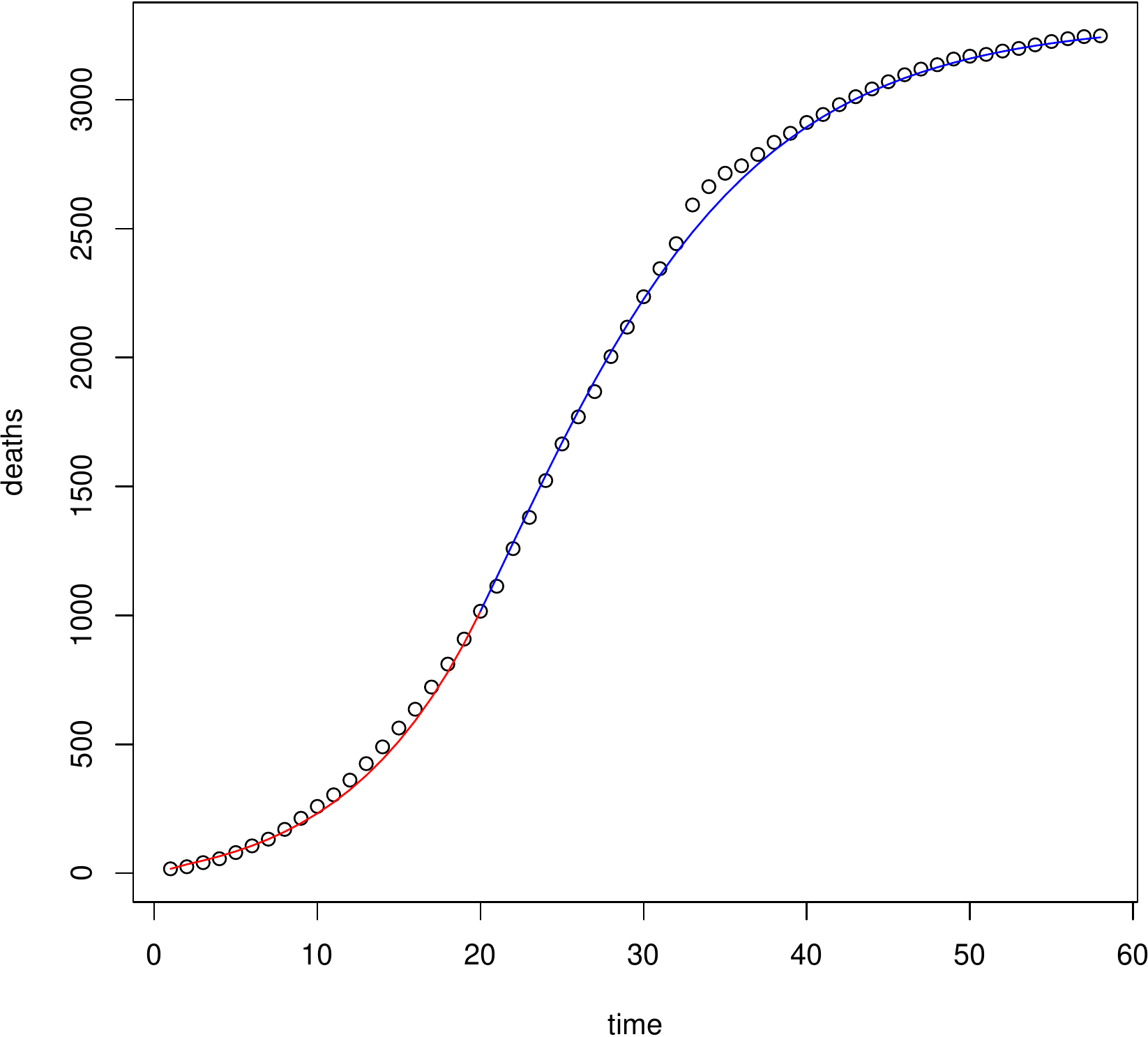}
 \caption{This plot shows the numbers of deaths in China from January 22 to March 19, together with the fitted function $D$.}\label{fig-china-2}
\end{figure}

Figure \ref{fig-china-2} shows the data points from Figure \ref{fig-china} together with a plot of the function $D$ with the fitted parameters $(\beta,\lambda)$. The red curve shows the function $D$ in the first period $[t_1,T_1]$, where $D$ proceeds convex, and the blue curve shows the function $D$ in the second period $[T_1,\infty)$, where $D$ proceeds concave and the numbers of deaths begin to converge. In this model January 24 is considered as the date of governmental measures, and hence the date $T_1$, where the curve changes from red to blue, is February 10.

\section{Estimates for data from Italy and other countries}\label{sec-Italy}

\begin{figure}[!ht]
 \centering
 \includegraphics[width=0.4\textwidth]{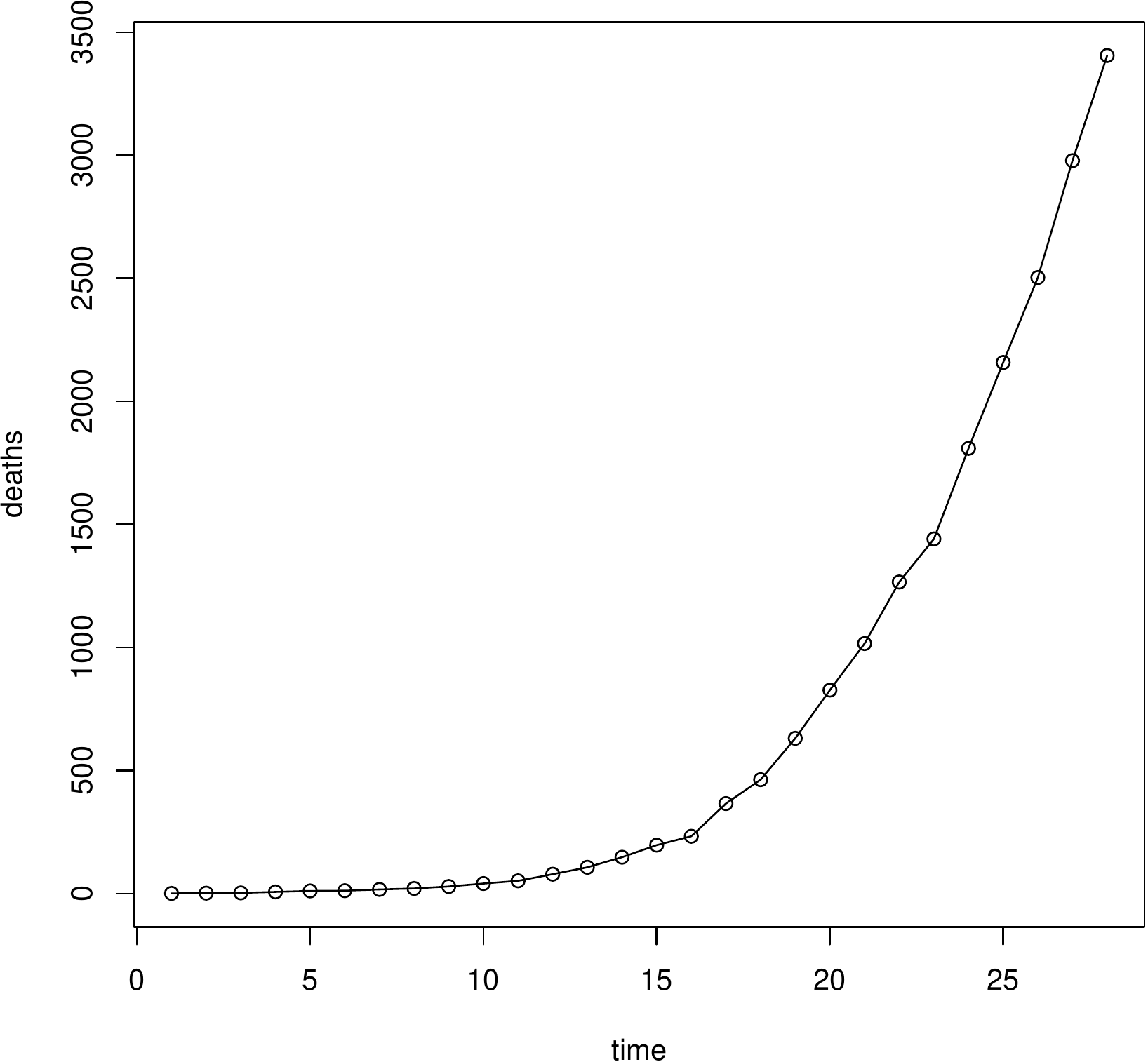}
 \includegraphics[width=0.4\textwidth]{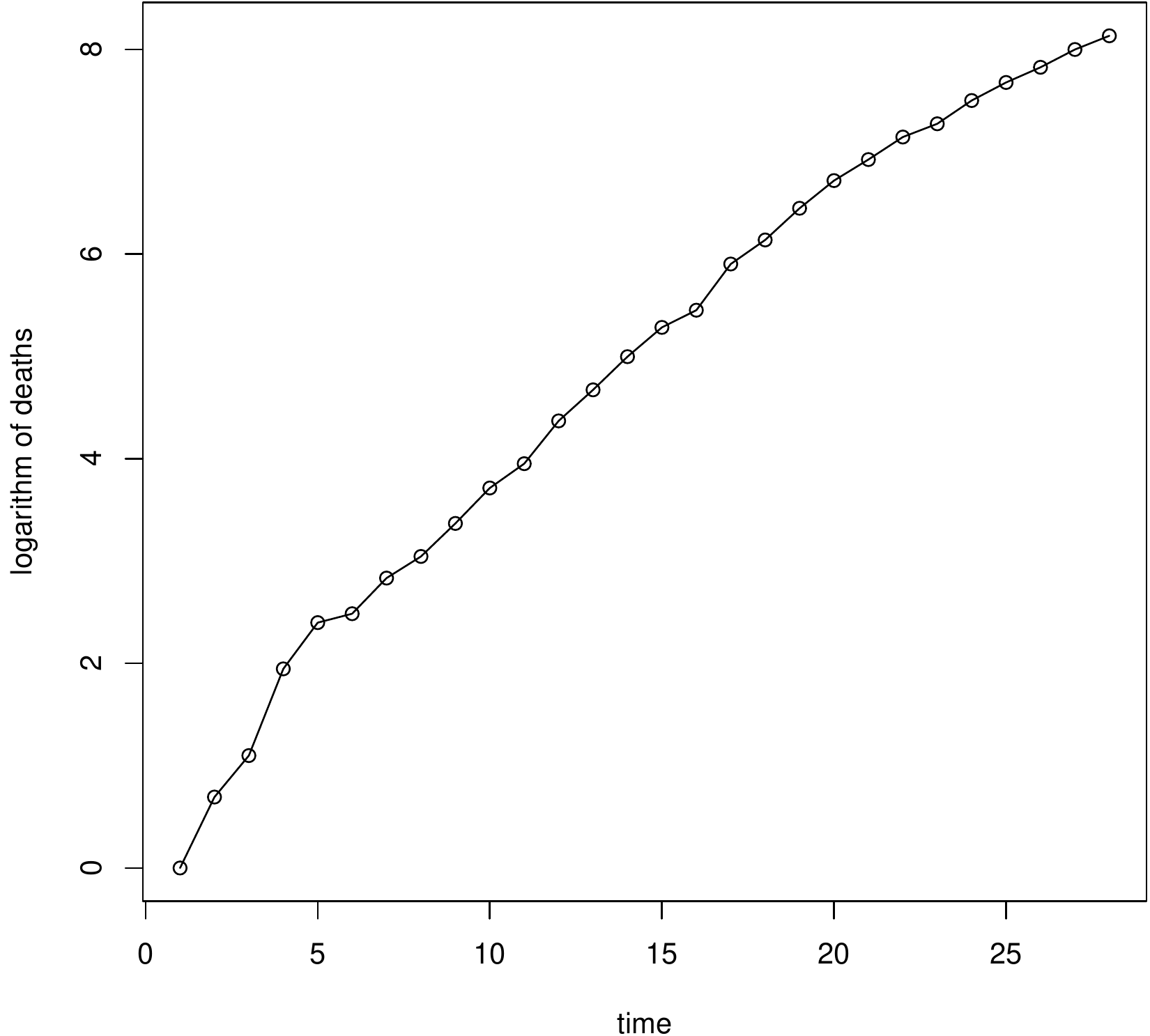}
 \caption{These two plots show the numbers of deaths in Italy from February 21 to March 19; in the left plot with the usual linear scale, and the right plot with logarithmic scale.}\label{fig-italy}
\end{figure}

As already pointed out, China is the only country which has essentially already overcome the corona crisis. South Korea has already reached the second stage of the corona crisis, and the numbers of deaths is comparatively low. Apart from that, the remaining countries are just in the first period, and we can make predictions of the numbers of deaths in the future. As an illustrating example, we consider the data from Italy. Figure \ref{fig-italy} shows the numbers of deaths due to the corona virus in Italy from February 21 to March 19, with linear and logarithmic scale.

At first view, the right plot in Figure \ref{fig-italy} looks linear, but actually we obtain the concavity parameter $\beta = 0.65$. Since Italy is in the first stage of the corona crisis, it is difficult to estimate the parameter $\lambda$ concerning the effects of governmental measures. However, since the measures are equally tight as those in China, it is reasonable to take the same parameter $\lambda = \ln(3.25)$.

\begin{figure}[!ht]
 \centering
 \includegraphics[width=0.5\textwidth]{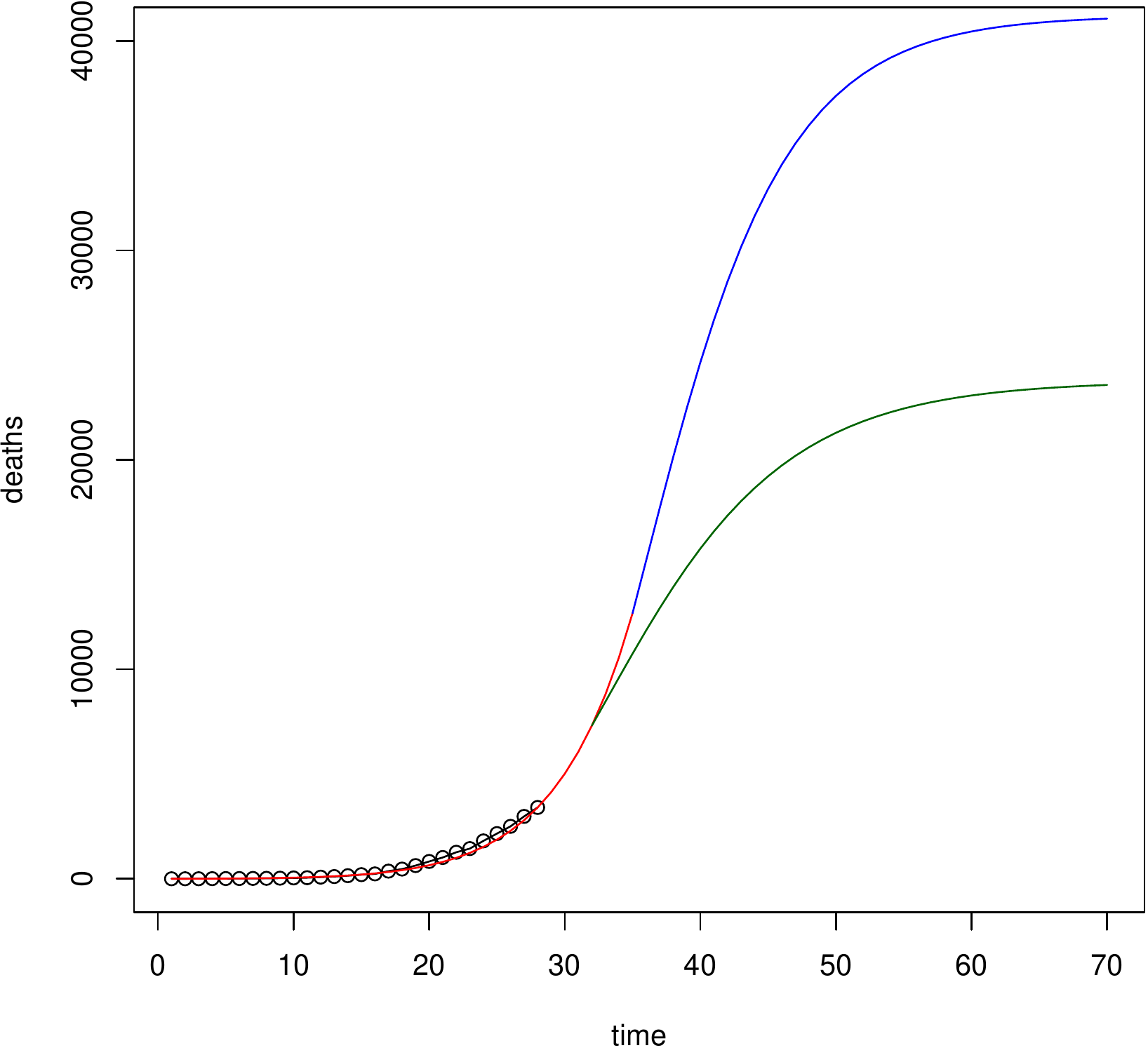}
 \caption{This plot shows a prediction of the numbers of deaths in Italy from February 21 to April 30, together with the fitted function $D$.}\label{fig-italy-3}
\end{figure}

Figure \ref{fig-italy-3} shows the available data points together with the predicted scenario according to the estimated parameters $(\beta,\lambda)$. The red curve shows the function $D$ in the first period $[t_1,T_1]$, and the blue curve shows the function $D$ in the second period $[T_1,\infty)$. In this model March 11 is considered as the date of governmental measures, and hence the date $T_1$, where the curve changes from red to blue, is March 28. Additionally, the green curve shows the corresponding scenario with the same parameters, but date $T_1$ being March 25. As a consequence, in this model governmental measures taking place three days earlier would bisect the number of deaths in the long run. Of course, other scenarios with different values of $\lambda$ can be considered as well.

We can perform the same procedure with other countries. For example, for Iran, Spain, Germany, USA, France, South Korea, Switzerland, United Kingdom, Netherlands and Belgium data are available. For most of the countries the concavity parameter is easy to fit; exceptions are USA and France (despite many data) as well as countries for which so far only few data exists. Apart from South Korea, for each country the concavity parameter $\beta$ is larger than $0.6$, which is the value for China. However, typically the concavity parameter is less than one, which means that the function $L$ is really concave, and not linear. Two exceptions are Germany and United Kingdom, where the current estimates are $\beta = 1$.

\end{document}